\documentclass[11pt]{article}
\usepackage{moriond,epsfig}

\def\Journal#1#2#3#4{{#1} {\bf #2}, #3 (#4)}

\def\NIMA{{\em Nucl. Instrum. Methods} A}
\def\NPB{{\em Nucl. Phys.} B}
\def\PLB{{\em Phys. Lett.}  B}
\def\PRL{\em Phys. Rev. Lett.}

\def\JHEP{\em J. High Energy Phys.}
\def\EPJ{{\em Eur. Phys. J.} C}

\def\be{\begin{equation}}
\def\ee{\end{equation}}
\def\bea{\begin{eqnarray}}
\def\eea{\end{eqnarray}}

\begin{document}
\vspace*{4cm}
\title{Beyond the Standard Model Higgs Boson Searches at the Tevatron}

\author{Tim Scanlon on behalf of the D0 and CDF Collaborations}

\address{Imperial College London, Physics Department,\\
Prince Consort Road, London SW7 2AZ, United Kingdom.}

\maketitle\abstracts{
Results are presented for beyond the Standard
Model Higgs boson searches using up to 8.2~fb$^{-1}$ of data from
Run II at the Tevatron. No significant excess is observed in any of
the channels so 95\% confidence level limits are presented.
}

\section{Introduction}

The search for the Higgs boson is one of the main goals in High Energy Physics and one of the highest priorities at Run II of the Tevatron. There are many alternative
Higgs boson models beyond the SM, including Supersymmetry
(SUSY)~\cite{mssm}, Hidden Valley (HV)~\cite{hv1,hv2} and Fermiophobic Higgs bosons~\cite{haber}, which can
actively be probed at the Tevatron, and in the absence of an excess
constrained. The latest limits for several SUSY Higgs boson searches are presented
in Section~\ref{sec:mssm}, for HV Higgs boson searches in Section~\ref{sec:hv} and for Fermiophobic Higgs boson searches in
Section~\ref{sec:fermiophobic}. More information on all
these searches, along with the latest results, can be found on the CDF and D0 public results webpages~\cite{cdf_web_page,dzero_web_page}.

\section{Minimal Supersymmetric Standard Model Higgs Boson Searches} \label{sec:mssm}

The Minimal Supersymmetric extension of the SM (MSSM) \cite{mssm}
introduces two Higgs doublets which results in five physical Higgs
bosons after electroweak symmetry breaking. Three of the Higgs bosons are
neutral, the CP-odd scalar, $A$, and the CP-even scalars, $h$ and
$H$ ($h$ is the lighter and SM like), and two are
charged, $H^{\pm}$.

At tree level only two free parameters are needed for all couplings
and masses to be calculated. These are chosen as the mass of the
CP-odd pseudoscalar ($m_{A}$) and tan$\beta$, the ratio of the two vacuum
expectation values of the Higgs doublets.

The Higgs boson production cross section in the MSSM is proportional to
the square of tan$\beta$. Large values of tan$\beta$ thus result in
significantly increased production cross sections compared to the
SM.  Moreover, one of the CP-even scalars and the CP-odd scalar are
degenerate in mass, leading to a further approximate doubling of the
cross section.

The main production mechanisms for the neutral Higgs bosons are the
$gg, b\bar{b} \rightarrow\phi$ and $gg, q\bar{q} \rightarrow \phi +
b\bar{b}$ processes, where $\phi = h, H, A$. The branching ratio of
$\phi \rightarrow b\bar{b}$ is around 90\% and $\phi \rightarrow
\tau^{+}\tau^{-}$ is around 10\%. This results in three channels of
interest: $\phi \rightarrow \tau^{+}\tau^{-}$, $\phi b\rightarrow
b\bar{b}b$ and $\phi b\rightarrow \tau^{+}\tau^{-}b$. The overall
experimental sensitivity of the three channels is similar due to the
lower background from the more unique signature of the $\tau$
decays.

\subsection{Higgs $\rightarrow \tau^{+}\tau^{-}$} \label{tautau}



D0's most recent search is in the $\tau_{\mu}\tau_{had}$ final state
using 1.2 fb$^{-1}$ of Run II data, where $\tau_{had}$ refers to a
hadronic decay and $\tau_{\mu}$ to a leptonic decay (to a $\mu$) of the $\tau$. This result is an extension to, and combined with, the published
1~fb$^{-1}$ result which also included the $\tau_{\mu}\tau_{e}$,
and $\tau_{e}\tau_{had}$ channels~\cite{p17tautau}. CDF have published a search combining the $\tau_{\mu}\tau_{e}$,
$\tau_{\mu}\tau_{had}$ and $\tau_{e}\tau_{had}$ final states using
1.8~fb$^{-1}$ of RunII data~\cite{cdftautau}.

Both searches require events to have an isolated $\mu$ ($e$), separated from an opposite signed $\tau_{had}$ (or $e$ for the $\tau_{\mu}\tau_{e}$ channel). Hadronic $\tau$
candidates are identified at D0 by neural
networks designed to distinguish $\tau_{had}$ from multi-jet events and at CDF by using a variable size isolation cone. To
minimise the $W$+jets background events are removed which have a large
$W$ transverse mass (D0) or by placing a cut on the relative
direction of the visible $\tau$ decay products and the missing
$E_{T}$ (CDF).

\begin{figure}[htb]
\centering
\includegraphics[width=0.42\columnwidth]{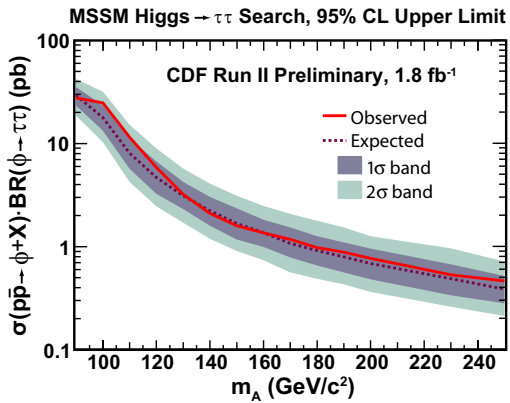}
\caption{Model independent $95\%$ CL upper limit on the branching ratio multiplied by cross section from the 1.8~fb$^{-1}$ CDF publication. The dark (light) grey bands show the 1 (2) standard deviation bands around the expected limit.
\label{Fig:tautaulimits}}
\end{figure}

In both analyses the $Z/\gamma \rightarrow \tau\tau$ and $W$+jets backgrounds are modelled using PYTHIA~\cite{pythia}, with the $W$+jets normalisation and the multi-jet contribution modelled using data. Limits are set using the visible mass distribution ($m_{vis}$),
which is the invariant mass of the visible $\tau$ products and the
missing $E_{T}$. The CDF model independent 95\% CL upper limit on the branching ratio multiplied by cross section is shown in
Fig.~\ref{Fig:tautaulimits}.

\subsection{Higgs $+~b\rightarrow b\bar{b}b$}


This channel has a signature of at least three $b$ jets, with the background consequentially dominated by heavy flavour multi-jet
events. D0 have recently published a search in this channel using 5.2~fb$^{-1}$ of data~\cite{hbbpub} and CDF have a preliminary result using 2.2~fb$^{-1}$ of data.

Both searches require three $b$-tagged jets, D0 uses its
standard neural network $b$-tagging algorithm~\cite{btagthesis} and CDF uses its standard secondary vertex algorithm. Due to the difficultly of simulating the heavy flavour multi-jet background both analyses use data driven approaches. D0 uses a fit to data over several different $b$-tagging criteria whereas CDF uses fits to dijet invariant and secondary vertex mass templates to determine the heavy flavour sample composition.

\begin{figure}[htb]
\centering
\includegraphics[width=0.38\columnwidth]{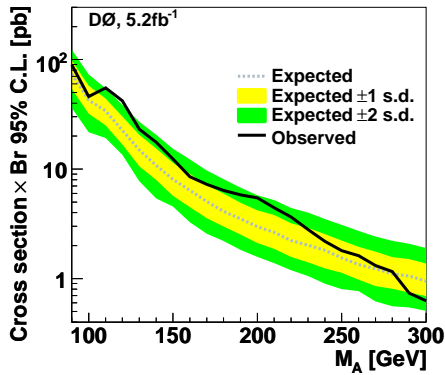}
\caption{
Model independent $95\%$ CL upper limit on the branching ratio multiplied by cross section from the 5.2~fb$^{-1}$ D0 publication. The yellow (green) bands show the 1 (2) standard deviation bands around the expected limit.
\label{Fig:hbblimits}}
\end{figure}

To increase the sensitivity of the analysis, D0 splits it into exclusive three
and four-jet channels, training a likelihood to distinguish the signal from background in each. Limits are set by both CDF and D0 on the Higgs boson production cross section
times branching ratio using the dijet invariant mass as the discriminating
variable. The D0 model independent $95\%$ CL upper limit on the branching ratio multiplied by cross section is shown in Fig.~\ref{Fig:hbblimits}.

\subsection{Higgs $+~b\rightarrow \tau^{+}\tau^{-}b$} \label{btautau}


D0 has performed a search for both the $\tau_{\mu}\tau_{had}$ and $\tau_{e}\tau_{had}$ signatures using 4.3~fb$^{-1}$ and 3.7~fb$^{-1}$ of Run II data respectively. Events are selected by requiring
an isolated muon or electron separated from an opposite sign
$\tau_{had}$ candidate, along with a $b$-tagged jet. The $\tau_{had}$ decays are
identified using the standard D0 neural networks and $b$ jets using the neural network
$b$-tagging algorithm. The dominant backgrounds are $t\bar{t}$, $W$+jets, multi-jet and $Z$+jet events. The
multi-jet and $W$+jets backgrounds are estimated from data with $t\bar{t}$
modelled using ALPGEN~\cite{alpgen} interfaced with PYTHIA.

\begin{figure}[htb]
\centering
\includegraphics[width=0.42\columnwidth]{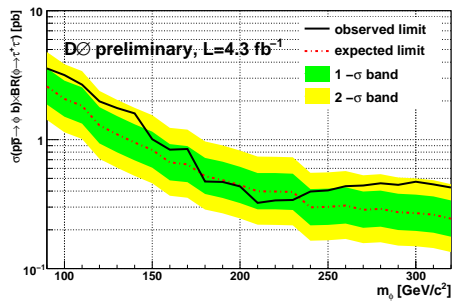}
\caption{Model independent $95\%$ CL upper limit on the branching ratio multiplied by cross section from the 4.2~fb$^{-1}$ D0 $\tau_{\mu}\tau_{had}$ channel. The green (yellow) bands show the 1 (2) standard deviation bands around the expected limit.}
\label{Fig:btautaulimits}
\end{figure}

To improve the sensitivity of the analysis discriminants are trained which differentiate the signal from the $t\bar{t}$, multi-jet and $Z$+light parton (muon channel only) events respectively. The discriminants are combined to form a final discriminant which is used to set limits.
Figure~\ref{Fig:btautaulimits} shows the model independent 95\% CL upper limit for the D0 $\tau_{\mu}\tau_{had}$ channel.

\subsection{Combined Limits}

\begin{figure}[h]
\centering
\includegraphics[width=0.38\columnwidth]{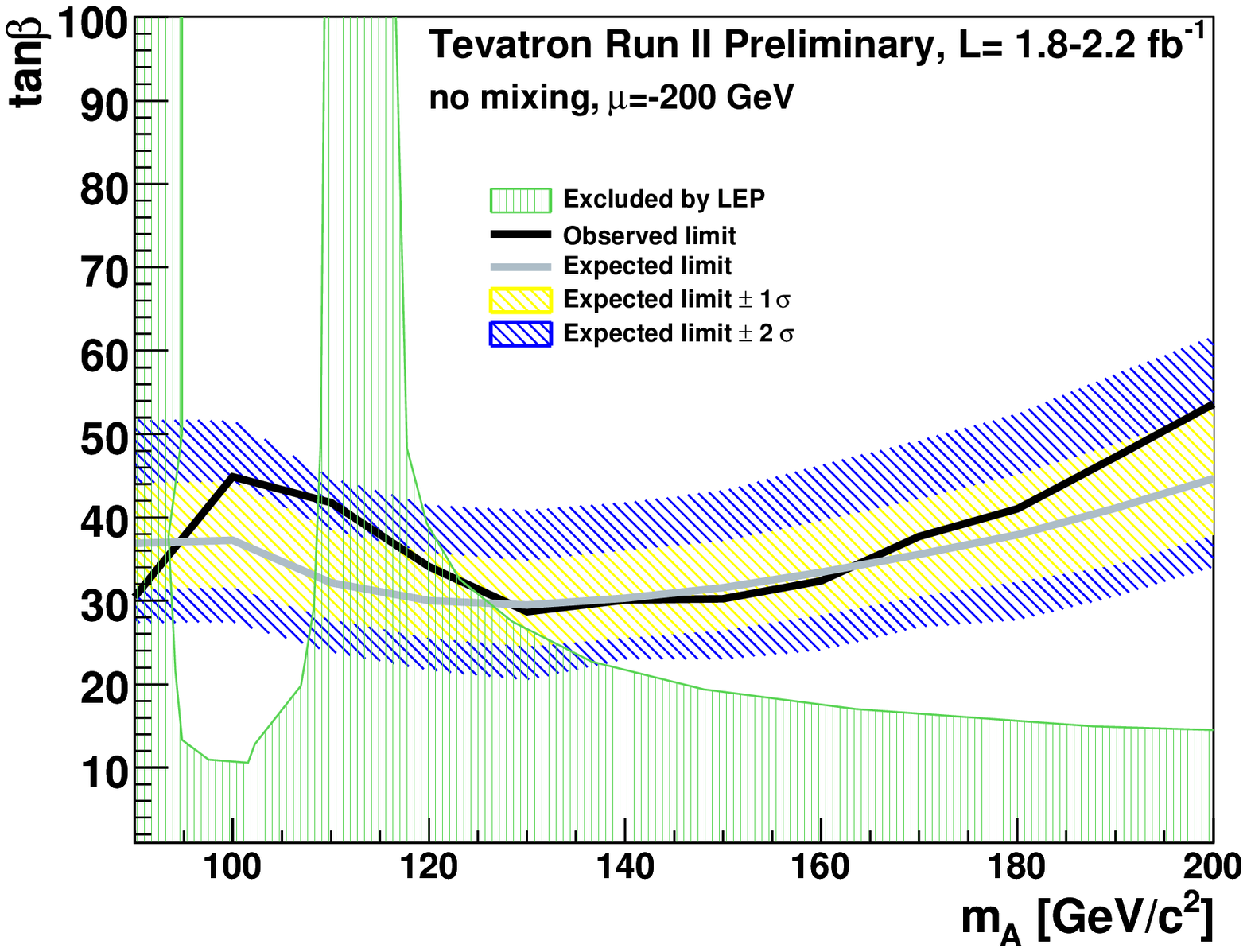}
\includegraphics[width=0.38\columnwidth]{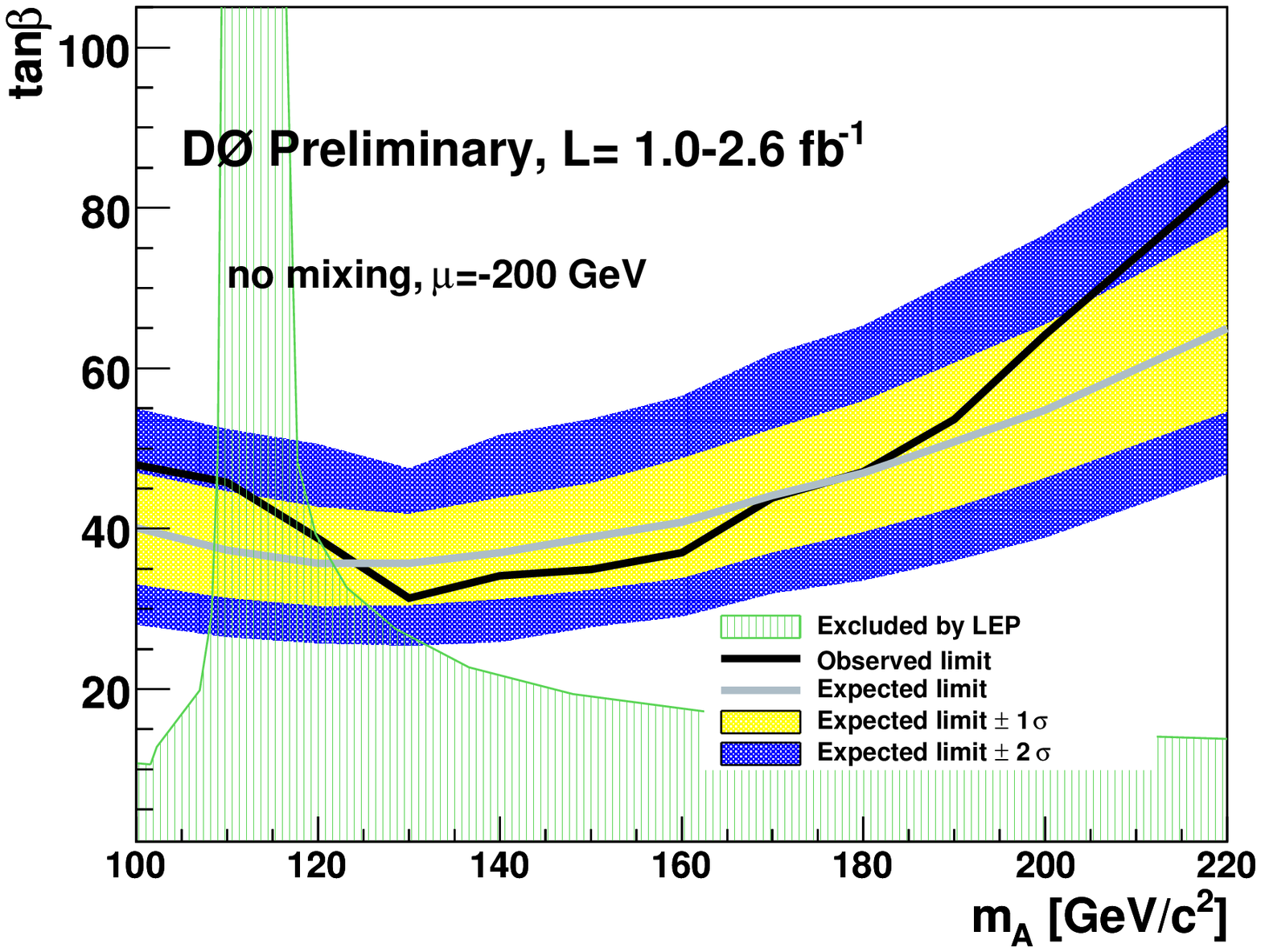}
\caption{
The combined D0 (left) and Tevatron (right) 95\% CL limits on tan$\beta$ versus m$_{a}$ for the $\mu<0$, no mixing scenario. The green area is the region excluded by LEP.  \label{Fig:tevcomblimits}}
\end{figure}

The channels described in Sections~\ref{tautau}--\ref{btautau} are complementary and can be combined to increase the reach of the MSSM Higgs boson searches at the Tevatron. D0 has combined its three neutral Higgs boson channels (using an earlier version of the $\tau^{+}\tau^{-}b$ analysis based on only 1.2~fb$^{-1}$ of RunII data and not including the $\tau_{e}\tau_{had}$ channel) and interpreted the limits in the standard MSSM scenarios~\cite{scenarios}. A combined Tevatron limit on the MSSM Higgs sector has also been produced from D0 and CDF's Higgs $\rightarrow \tau^{+}\tau^{-}$ channels. The combined Higgs $\rightarrow \tau^{+}\tau^{-}$ result has been interpreted in a quasi-model independent limit, as well as in the standard scenarios. Both the D0 and Tevatron combined 95\% CL limits for one of the scenarios are shown in Fig.~\ref{Fig:tevcomblimits} along with the limit from LEP~\cite{lep}.

\subsection{Next-to-MSSM Higgs Bosons Searches}


In the next-to-MSSM (nMSSM)~\cite{nmssm} the branching ratio of
Higgs$\rightarrow b\bar{b}$ is greatly reduced. Instead the Higgs
boson predominantly decays to a pair of lighter neutral pseudoscalor Higgs
bosons, $a$. The nMSSM scheme is interesting as it allows the LEP
limit on the $h$ boson to be naturally lowered to the general Higgs
boson search limit from LEP of $M_{h} > 82$~GeV~\cite{lepgeneral}.

\begin{figure}[htb]
\centering
\includegraphics[width=0.42\columnwidth]{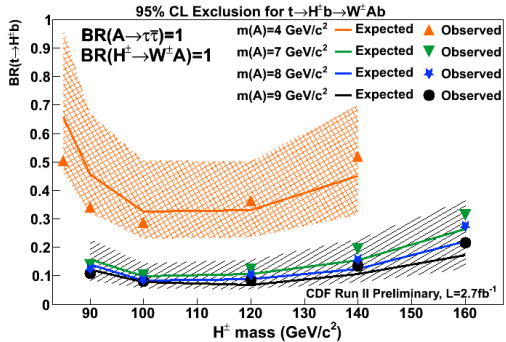}
\caption{The 95\% CL upper limits on branching ratio of top decaying to $H^{+}b$ for various Higgs bosons masses.}
\label{Fig:atautau}
\end{figure}

CDF has conducted a search for a light nMSSM Higgs boson using 2.7~fb$^{-1}$ of data in top quark decays, where $t\rightarrow W^{\pm(*)}ab$ and $a\rightarrow\tau\tau$. The $\tau$ particles are identified by the presence of additional isolated tracks in the event due to their low $p_{T}$. The dominant background is from soft parton interactions and is modelled using data. Upper limits are set at the 95\% CL on the branching ratio of a top quark decaying to a charged Higgs boson from a fit to the $p_{T}$ spectrum of the lead isolated track and are shown in Fig.~\ref{Fig:atautau}.

\section{Hidden Valley Higgs} \label{sec:hv}

CDF has conducted Higgs boson searches in Hidden Valley (HV) models, which contain long-lived particles which travel a macroscopic distance before decaying into two jets. The signature of this search is a Higgs boson decaying to two HV particles, which travel for $\sim1$~cm before decaying to two b-quarks. Although there are four b-jets present in the decay, to increase the efficiency only three are required, two of which must be $b$-tagged and not back-to-back in the detector. A specially adapted version of CDF's secondary vertex $b$-tagging tool is used to reconstruct the displaced secondary vertices and the reconstructed HV decay points are required to have a large decay length.

\begin{figure}[htb]
\centering
\includegraphics[width=0.42\columnwidth]{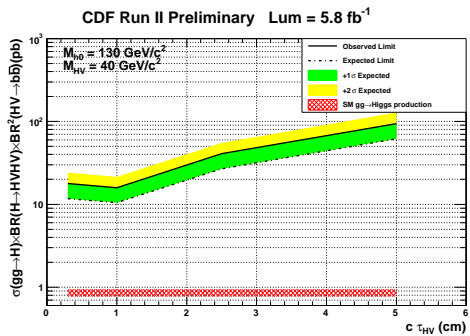}
\caption{The 95\% CL upper limit on $\sigma \times BR$ as a function of the
Hidden Valley particle's lifetime.}
\label{Fig:hiddenvalley}
\end{figure}

Due to the difficulty of usinge Monte Carlo to model background events with large decay lengths, a data driven approach is used. The predicted number of background events is compared to the number seen in data and in the absence of a significant excess, limits are set on the production cross section times branching ratio on the benchmark HV model. Figure~\ref{Fig:hiddenvalley} shows the 95\% CL upper limit for a Higgs mass of 130~GeV and a HV particle mass of 40~GeV as a function of the HV particle lifetime.

\section{Fermiophobic Higgs Boson Searches}
\label{sec:fermiophobic}

The Standard Model Higgs boson branching ratio to a pair of
photons is small. There are however several models where the decay
of the Higgs boson to fermions is suppressed. In these models the
decay of the Higgs boson to photons is greatly enhanced. Both D0
and CDF \cite{cdffermiphobic} have carried out searches for the Fermiophobic Higgs boson
using 8.2 and 4.2 fb$^{-1}$ respectively.

D0 requires two photon candidates in the central calorimeter, with jets misidentified as photons rejected by use of a neutral network. Electrons are suppressed by requiring
that the photon candidates are not matched to activity in the
tracking detectors. A decision tree is trained using five variables to distinguish signal from background events. The three main background sources are estimated
separately: the jet and diphoton backgrounds are estimated from data
and the Drell-Yan contribution is estimated using PYTHIA.

CDF's search also requires two photons, with only one of them required to be in the central region of the calorimeter. This looser photon requirement approximately doubles the acceptance compared to requiring both photons in the central region. In addition a cut is placed on the transverse
momentum of the two photons which significantly reduces the background, which is estimated using
a purely data-based approach.

\begin{figure}[htb]
\centering
\includegraphics[width=0.42\columnwidth]{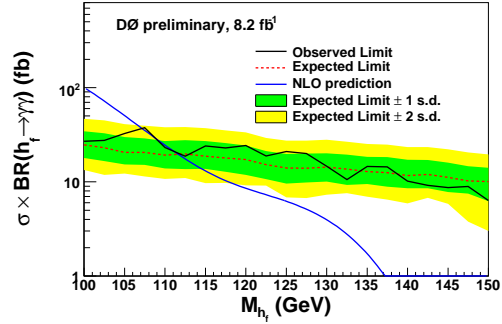}
\caption{The 95\% CL upper limit on $\sigma \times BR$ as a function of the
Fermiophobic Higgs boson mass for D0.}
\label{Fig:fermiophobichiggs}
\end{figure}

Upper limits are set on the Higgs boson production cross section times
branching ratio using the decision tree output (D0) or diphoton mass (CDF) as the discriminating
variable. The 95\% CL upper limit are shown in Fig.~\ref{Fig:fermiophobichiggs}
for the D0 search.


\section{Conclusions}

CDF and D0 have a wide variety of beyond the Standard Model
Higgs boson searches, presented here using up to 8.2~fb$^{-1}$ of data. These
searches are already powerful, and have set some of the best limits in
the world. No signal has been observed yet, but with their rapidly improving sensitivity, due to both improved analysis techniques and the addition of between 2--5 times more data (which has already been recorded), these analyses will continue to probe extremely interesting regions of parameter space, promising many exciting results in the near future.

\section*{Acknowledgments}
I would like to thank all the staff at Fermilab, the Tevatron accelerator division along with the CDF and the D0 Collaborations.

\end{document}